\documentclass[12pt,preprint]{aastex}
\usepackage{graphicx,subfigure,amsmath}
\newcommand\gsim{\,\lower3pt\hbox{$\sim$}\llap{\raise2pt\hbox{$>$}}\,}
\newcommand\lsim{\,\lower3pt\hbox{$\sim$}\llap{\raise2pt\hbox{$<$}}\,}

\begin{document}
\title{An MHD Model of the December 13 2006 Eruptive Flare}

\author{Y.~Fan}
\affil{High Altitude Observatory, National Center for Atmospheric
Research\altaffilmark{1}, 3080 Center Green Drive, Boulder, CO 80301}

\altaffiltext{1} {The National Center for Atmospheric Research
is sponsored by the National Science Foundation}

\begin{abstract}
We present a 3D MHD simulation that qualitatively models the coronal magnetic
field evolution associated with the eruptive flare that occurred on December 13,
2006 in the emerging $\delta$-sunspot region NOAA 10930 observed by the Hinode
satellite. The simulation is set up where we drive the emergence of an
east-west oriented magnetic flux rope at the lower boundary into a
pre-existing coronal field constructed from the SOHO/MDI
full-disk magnetogram at 20:51:01 UT on
December 12, 2006. The resulting coronal flux rope embedded in the ambient
coronal magnetic field first settles into a stage of quasi-static rise, and
then undergoes a dynamic eruption, with the leading edge of the flux rope cavity
accelerating to a steady speed of about 830 km/s.
The pre-eruption coronal magnetic field shows morphology that is in
qualitative agreement with that seen in the Hinode soft X-ray observation in
both the magnetic connectivity as well as the development of an inverse-S
shaped X-ray sigmoid. We examine the properties of the erupting flux rope and
the morphology of the post-reconnection loops, and compare them with the
observations.
\end{abstract}

\section{Introduction}
Coronal mass ejections (CMEs) are large-scale, spontaneous ejections of plasma
and magnetic flux from the lower solar corona into interplanetary space and
are major drivers of space weather near
earth \citep[e.g.][]{hundhausen1993, lindseyetal1999, webbetal2000}.
CMEs and eruptive flares are believed to result from a sudden, explosive
release of the free magnetic energy stored in the previously
quasi-equilibrium, twisted/sheared coronal magnetic
field \citep[see e.g. reviews by][]{forbesetal_2006,chen2011}.
Using idealized constructions, both analytical studies and numerical
simulations have been carried out to understand the basic underlying 
magnetic field structures of the eruption precursors, and
the physical mechanisms of their sudden eruption
\citep[e.g.][]{mikic_linker1994,antiochosetal1999,forbes_priest1995,
linetal1998,amarietal2000,sturrock2001,roussevetal2003,
toeroek_kliem2005,toeroek_kliem2007,fan_gibson2007,isen_forbes2007,fan2010,
aulanieretal2010,demoulin_aulanier2010}.
Magneto-hydrodynamic (MHD) models of observed CME events have also been
constructed
to determine the actual magnetic field evolution and causes for
the eruption and the properties of the magnetic ejecta, which are critical
for determining the geo-effectiveness of the resulting
interplanetary coronal mass ejections (ICMEs) \citep[e.g.][]{mikicetal2008,
titovetal2008,kataokaetal2009}.

The eruptive event in active region 10930 on December 13, 2006 produced
an X3.4 flare and a fast, earth-directed CME with an estimated speed of at
least 1774 km/s.
The ICME reached the Earth on 14-15 December 2006, with a strong and
prolonged southward directed magnetic field in the magnetic cloud, causing
a major geomagnetic storm \citep[e.g.][]{liuetal2008,kataokaetal2009}.
This event is particularly well observed by Hinode for both the coronal
evolution as well as the photospheric magnetic field evolution over a period
of several days preceding, during, and after the eruption.
The photospheric magnetic field evolution of AR 10930 was characterized by
an emerging $\delta$-sunspot with a growing positive polarity, which displayed
substantial (counter-clockwise) rotation and eastward motion as it
grew (see e.g. the movie provided at the NOAJ website
http://solar-b.nao.ac.jp/news/070321Flare/me\_20061208\_15arrow\_6fps.mpg and
see also \citet{min_chae2009}).
This is indicative of the emergence of a
twisted magnetic flux rope with the positive rotating spot being one of its
photospheric footpoints.
The total rotation of the positive, growing sunspot prior to the onset
of the flare is measured to be $240^{\circ}$ by \citet{zhangetal2007}
and $540^{\circ}$ by \citet{min_chae2009}, which gives an estimate of the
minimum amount of twist that has been transported into the corona in the
emerged flux rope.

Several studies based on non-linear force-free field extrapolations from the
photospheric vector magnetic field measurement for AR 10930 have been
carried out to study the coronal magnetic field and the associated free
magnetic energy before and after the flare \citep[e.g.][]{schrijveretal2008,
inoueetal2008}.
In this paper, we present an MHD simulation that model the coronal magnetic
field evolution associated with the onset of the eruptive flare in AR 10930
on December 13 2006.
The simulation assumes the emergence of an east-west oriented
magnetic flux rope into a pre-existing coronal magnetic field constructed
based on the SOHO/MDI full-disk magnetogram of the photospheric magnetic field
at 20:51:01 UT on December 12.  
Our simulated coronal magnetic field first achieves a quasi-equilibrium phase
during which the coronal flux rope rises quasi-statically as more twisted
flux is being transported into the corona through a slow flux emergence.
The evolution is then followed by a dynamic eruption, where the erupting flux
rope accelerate to a final steady speed of about 830 km/s.
The erupting flux rope is found to undergo substantial writhing or rotational
motion, and the erupting trajectory is non-radial, being deflected southward
and eastward from the local radial direction of the source region.
The coronal magnetic field structure just prior to the onset of the eruption
reproduces qualitatively the observed morphology and connectivity of
the coronal magnetic field, including the formation of an inverse-S
shaped pre-eruption sigmoid, as seen in the Hinode XRT images.
After the onset of the eruption, the evolution of the post-reconnection
loops and their foot-points resulting from the simulated magnetic field is
also in qualitative agreement with the morphology of the observed X-ray
post-flare brightening and the evolution of the chromosphere flare ribbons.

We organize the remainder of the paper as follows. In Section \ref{sec:model},
we describe the MHD numerical model and how the simulation is set up.  In
Section \ref{sec:result} we describe the resulting evolution of the simulated
coronal magnetic field and compare with observations.
We summarize the conclusions and discuss future directions for improving the
model in Section \ref{sec:conc}.

\section{Model Description\label{sec:model}}
For the simulation carried out in this study, we solve the following
magneto-hydrodynamic equations in a spherical domain:
\begin{equation}
\frac{\partial \rho}{\partial t}
+ \nabla \cdot ( \rho {\bf v}) = 0 ,
\label{eqcont}
\end{equation}
\begin{equation}
\rho \left ( \frac{\partial {\bf v}}{dt}
+ ({\bf v} \cdot \nabla ) {\bf v} \right )
= - \nabla p - \rho \frac{G M_{\odot}}{r^2} \hat{\bf r}+ \frac{1}{4 \pi}
( \nabla \times {\bf B} ) \times {\bf B},
\label{eqmotion}
\end{equation}
\begin{equation}
\frac{\partial {\bf B}}{\partial t}
= \nabla \times ({\bf v} \times {\bf B}),
\label{eqinduc}
\end{equation}
\begin{equation}
\nabla \cdot {\bf B} = 0,
\label{eqdivb}
\end{equation}
\begin{equation}
\frac{\partial e}{\partial t} = - \nabla \cdot
\left [ \left ( \varepsilon + \rho \frac{v^2}{2} + p \right ) {\bf v}
- \frac{1}{4 \pi} ({\bf v} \times {\bf B} ) \times {\bf B} \right ]
- \rho {\bf v} \cdot \frac{G M_{\odot}}{r^2} \hat{\bf r} ,
\label{eqetot}
\end{equation}
\begin{equation}
p = \frac{\rho R T}{\mu},
\label{eqstate}
\end{equation}
where
\begin{equation}
\varepsilon = {p \over {\gamma - 1} } .
\end{equation}
\begin{equation}
e=\varepsilon + \rho \frac{v^2}{2} + \frac{B^2}{8 \pi} .
\end{equation}
In the above ${\bf v}$, ${\bf B}$, $\rho$, $p$, $T$, $\varepsilon$, $e$
$R$, $\mu$, $\gamma$, $G$, and $M_{\odot}$ denote
respectively the velocity field, the magnetic field, density,
pressure, temperature, the internal energy density, the total
energy density (internal+kinetic+magnetic), the gas constant,
the mean molecular weight, the ratio of specific heats, the gravitational
constant, and the mass of the Sun. We have assumed an ideal polytropic
gas with $\gamma = 1.1$ for the corona plasma.
The above MHD equations are solved numerically without
any {\it explicit} viscosity, magnetic diffusion, and non-adiabatic effects.
However numerical dissipations are present, and since we are solving the
total energy equation in conservative form, the numerical dissipation of
kinetic, and magnetic energy is effectively being put back
into the internal energy.

The basic numerical schemes we use to solve the above MHD
equations are as follows.  The equations are discretized in
spherical domain with $r$, $\theta$, $\phi$ coordinates using a
staggered finite-difference scheme \citep{stone_norman1992a}, and
advanced in time with an explicit, second order accurate, two-step
predictor-corrector time stepping.
A modified, second order accurate Lax-Friedrichs scheme similar
to that described in \citet[][see eq. (A3) in that paper]
{rempeletal2008} is applied for evaluating
the fluxes in the continuity and energy equations.
Compared to the standard second order Lax-Friedrichs scheme,
this scheme significantly reduces numerical diffusivity for
regions of smooth variation, while retaining the same
robustness in regions of shocks.
The standard second order Lax-Friedrichs scheme is used for evaluating
the fluxes in the momentum equation.
A method of characteristics that is upwind in the Alfv\'en
waves \citep{stone_norman1992b} is used for evaluating the
${\bf v} \times {\bf B}$ term in the induction equation, and
the constrained transport scheme is used to ensure $\nabla 
\cdot {\bf B} = 0$ to the machine precision.

The simulation is set up where we drive the emergence of a part of a twisted
magnetic torus at the lower boundary into a pre-existing coronal potential
field, constructed based on the MDI full-disk magnetogram from 20:51:01 UT on
December 12, 2006 (Figure \ref{fig1}a).
First, from the full-disk MDI magnetogram, a region centered on the
$\delta$-spot (the white box in Figure \ref{fig1}a), with an latitudinal extent
of $30^{\circ}$ and a longitudinal extent of $45^{\circ}$ is extracted as
the lower boundary of the spherical simulation domain. In terms of the
the simulation coordinates, the domain
spans $r \in [R_{\odot}, 6.25 R_{\odot}]$,
$\theta \in [75^{\circ}, 105^{\circ}]$,
$\phi \in [-22.5^{\circ},22.5^{\circ}]$, with the center of
its lower boundary: $\theta= 90^{\circ}$ and $\phi=0^{\circ}$,
corresponding to the
center of the white-boxed area in Figure \ref{fig1}a.
This domain is resolved by a
grid of $512 \times 352 \times 528$, with 512 grid points in $r$, 352 grid
points in $\theta$, and 528 grid points in $\phi$.  The grid is uniform in
the $\theta$ and $\phi$ directions but non-uniform in $r$, with a uniform
grid spacing of $dr = 1.028$ Mm in the range of $r= R_{\odot}$ to
about $1.6 R_{\odot}$ and a geometrically increasing grid spacing above
$1.6 R_{\odot}$, reaching about $dr = 173.4$ Mm at the outer boundary.
We assume perfectly conducting walls for the side boundaries, and for the
outer boundary we use a simple outward extrapolating boundary condition that
allows plasma and magnetic field to flow through.
The lower boundary region extracted from the MDI full disk magnetogram
(as viewed straight-on) is shown in Figure \ref{fig1}b,
where we simply take the interpolated line-of-sight flux density from the
full-disk magnetogram and assume that the magnetic field is normal
to the surface to obtain the $B_r$ shown in the Figure.
The region contains roughly all the flux of the $\delta$-spot and the
surrounding pores and plages, to which some of the flux of the $\delta$-spot
is connected. 
The peak field strength in the region is about $3000$ G.
A smoothing using a Gaussian filter is carried out on the lower boundary region
until the peak field strength is reduced to about $200$ G.
This is necessary since the simulation domain corresponds to the corona, with
the lower boundary density assumed to be that of the base of the corona, and
thus a significant reduction of the field strength from that measured on the
photosphere is needed to avoid unreasonably high Alf\'ven speeds, which would
put too severe a limit on the time step of numerical integration.
After the smoothing, the magnetic flux in a central area, which roughly
encompasses the region of the observed flux emergence (including the rotating,
positive sunspot) is zeroed out (see Figure \ref{fig1}c) to be the area where
the emergence of an idealized, twisted magnetic torus is driven on the lower
boundary.
The potential field constructed from this lower boundary normal flux
distribution in Figure \ref{fig1}c is assumed to be the initial
coronal magnetic field for our simulation, which is shown in Figure \ref{fig2}.
We zero out the normal flux in the area for driving the flux emergence
so that we can specify analytically the subsurface emergence structure
in a field free region without the complication of the subsurface extension
of a pre-existing flux in the same area.

The initial atmosphere in the domain is assumed to be a static
polytropic gas:
\begin{equation}
\rho = \rho_0 \left [ 1 - \left ( 1- \frac{1}{\gamma} \right )
\frac{GM_{\odot}}{R_{\odot}} \frac{\rho_0}{p_0} \left ( 1
- \frac{R_{\odot}}{r} \right ) \right ]^{\frac{1}{1-\gamma}}
\end{equation}
\begin{equation}
p = p_0 \left [ 1 - \left ( 1- \frac{1}{\gamma} \right )
\frac{GM_{\odot}}{R_{\odot}} \frac{\rho_0}{p_0} \left ( 1
- \frac{R_{\odot}}{r} \right ) \right ]^{\frac{\gamma}{1-\gamma}}
\end{equation}
where $\rho_0 = 8.365 \times 10^{-16} $ g ${\rm cm}^{-3}$,
and $p_0 = 0.152$ dyne ${\rm cm}^{-2}$ are respectively the density and
pressure at the lower boundary of the coronal domain, and the
corresponding assumed temperature at the lower boundary is 1.1 MK.
The initial magnetic field in the domain is potential, and thus does not
exert any forcing on the atmosphere which is in hydrostatic equilibrium.
Figure \ref{fig3} shows the height profiles of the Alfv\'en speed and the
sound speed along a vertical line rooted in the peak $B_r$ of the main
pre-existing negative polarity spot.
For the initial state constructed, the peak
Alfv\'en speed is about 24 Mm/s, and the sound speed is 141 km/s at the
bottom and gradually declines with height. In most of the simulation domain,
the Alfv\'en speed is significantly greater than the sound speed.

At the lower boundary (at $r=R_{\odot}$), we impose (kinematically) the
emergence of a twisted torus ${\bf B}_{\rm tube}$ by specifying
a time dependent transverse electric field
${\bf E}_{\perp}|_{r=R_{\odot}}$ that corresponds to the upward advection
of the torus with a velocity ${\bf v}_{\rm rise}$:
\begin{equation}
{\bf E}_{\perp}|_{r=R_{\odot}} = {\hat{\bf r}} \times \left [ \left (
- \frac{1}{c} \, {\bf v}_{\rm rise} \times
{\bf B}_{\rm tube} (R_{\odot}, \theta, \phi, t) \right )
\times {\hat{\bf r}} \right ].
\label{eq_emf}
\end{equation}
The magnetic field ${\bf B}_{\rm tube}$ used for specifying
${\bf E}_{\perp}|_{r=R_{\odot}}$ is an axisymmetric torus
defined in its own local spherical polar coordinate system
($r'$, $\theta'$, $\phi'$) whose polar axis is the symmetric axis
of the torus. In the sun-centered simulation spherical
coordinate system, the origin of the ($r'$, $\theta'$, $\phi'$)
system is located at ${\bf r} = {\bf r}_c = (r_c, \theta_c, \phi_c)$,
and its polar axis (the symmetric axis of the torus) is in the plane of the
${\hat {\bf \theta}}$ and ${\hat {\bf \phi}}$ vectors at position ${\bf r}_c$
and tilted from the $- {\hat {\bf \theta}}$ direction clockwise
(towards the ${\hat {\bf \phi}}$ direction) by an angle $\delta$.
In the ($r'$, $\theta'$, $\phi'$) system,
\begin{equation}
{\bf B}_{\rm tube} = \nabla \times \left (
\frac{A(r',\theta')}{r' \sin \theta' } \hat{\bf \phi'} \right )
+ B_{\phi'} (r', \theta') \hat{\bf \phi'},
\end{equation}
where
\begin{equation}
A(r',\theta') = \frac{1}{4} q a^2 B_t
\left( 1 - \frac{\varpi^2(r',\theta')}{a^2} \right)^2 ,
\label{eqafunc}
\end{equation}
\begin{equation}
B_{\phi'} (r', \theta') = \frac{a B_t}{r' \sin \theta'}
\left( 1 - \frac{\varpi^2(r',\theta')}{a^2} \right).
\label{eqbph}
\end{equation}
In the above, $a$ is the minor radius of the torus,
$\varpi = (r'^2 + R'^2 -2r'R' \sin \theta')^{1/2}$ is
the distance to the curved axis of the torus, where $R'$ is the major
radius of the torus, $q$ denotes the angular amount (in rad) of field
line rotation about the axis over a distance $a$ along the axis, and $B_t a/R'$
gives the field strength at the curved axis of the torus.
The magnetic field ${\bf B}_{\rm tube}$ is truncated to zero outside of the
flux surface whose distance to the torus axis is $\varpi = a$.
We use $a = 0.035 R_{\odot}$, $R' = 0.063 R_{\odot}$,
$q/a = - 0.0308$ rad ${\rm Mm}^{-1}$, $B_t a/R' = 111$ G.
The torus center is assumed to be initially located at
${\bf r}_c = (r_c = 0.902 R_{\odot}, \, \theta_c = 90^{\circ}, \,
\phi_c = 0^{\circ} ) $, and the tilt of the torus $\delta=0$.
Thus the torus is initially entirely below the
lower boundary and is in the azimuthal plane.
For specifying ${\bf E}_{\perp}|_{r=R_{\odot}}$,
we assume that the torus moves bodily towards the lower boundary at a
velocity ${\bf v}_{\rm rise} = v_{\rm rise} \hat{{\bf r}}_c$,
where $v_{\rm rise}$ is described later.
The imposed velocity field at the lower boundary is a constant
${\bf v}_{\rm rise}$ in the area where the emerging torus intersects the
lower boundary and zero in the rest of the area.
The resulting normal flux distribution on the lower boundary after the imposed
emergence has stopped is shown
in Figure \ref{fig1}d. In it an east-west oriented bipolar pair has emerged,
where the positive spot represents the emerging, rotating positive
sunspot at the south edge of the dominant negative spot in Figure \ref{fig1}b,
and the negative spot corresponds to the flux in the fragmented
pores and plages to the west of the rotating positive
sunspot in Figure \ref{fig1}b. Observational
study by \citet{min_chae2009} found that the minor, fragmented pores of
negative polarity emerged and moved westward while the positive rotating
sunspot moved eastward, suggesting that they are the counterpart to which the
positive rotating sunspot is {\it at least partly} connected to (see Figure 2
in \citet{min_chae2009}).
This is one of the reasons that we model the coronal magnetic field in this
study with the emergence of an east-west oriented twisted flux rope.
After the emergence is stopped, the transverse electric field on the lower
boundary (eq. [\ref{eq_emf}]) is set to zero and the magnetic field is
line-tied at the lower boundary.
At the end of the emergence, the peak normal field strength in the emerged
bipolar region on the lower boundary reaches 121 G, compared to the 178 G peak
normal field strength in the dominant negative pre-existing spot in the
initial lower boundary field.
Due to the substantial smoothing of the observed normal magnetic flux density,
the total unsigned flux on the lower boundary of our
simulation is only about 30\% of that on the photosphere in the boxed area
shown in Figure \ref{fig1}a.
However, the ratio of the emerged flux (in the flux rope) over the total flux
on the lower boundary, $\sim 10$\%, for our simulation is about the same as
the ratio of the observed emerged flux (in the positive rotating sunspot) over
the total flux in the boxed area in Figure \ref{fig1}a.

Note that although the coronal temperature and density are used at the lower
boundary, the dynamic property of the lower boundary reflects the
property of the photosphere. The lower boundary is assumed to be ``infinitely
heavy'' such that the magnetic stress exerted on it from the corona does not
result in any motion of the field line foot-points (field anchoring or
line-tying) and that the lower boundary evolves in a prescribed way by a
kinematically imposed flux emergence associated with the upward advection of
a twisted flux rope.  Thus dynamically the lower boundary is meant to approximate
the photosphere, which can support cross-field currents and the resulting magnetic stresses.
However the thermodynamic conditions of the corona (instead of the
photosphere) are used for the lower boundary so that (1) we do not have to resolve
the small (about 150 km) photospheric pressure scale height in a simulation of the large
scale coronal evolution of a CME (size scale on the order of a solar radius),
and (2) we avoid solving the complex energy transport associated with coronal heating,
radiative cooling, and thermal conduction, which would be required if we were to include
the thermodynamics of the photosphere-chromosphere-corona system in the simulation.
Here for modeling the large scale, magnetically dominated dynamic evolution of the CME
initiation, we greatly simplify the thermodynamics (assuming an ideal polytropic gas
for the coronal plasma throughout the domain), and focus on the magnetic field evolution
of the corona in response to the imposed flux emergence and field-line anchoring representative
of the heavy photospheric lower boundary.

In the remainder of the paper, quantities are expressed in the following units
unless otherwise specified: $R_{\odot} = 6.96 \times 10^{10} \, {\rm cm}$,
$\rho_0 = 8.365 \times 10^{-16} \, {\rm g/cm^3}$, $B_0 = 20 \,
{\rm G}$, $v_{a0} = B_0 / \sqrt{4 \pi \rho_0 } = 1951 \,
{\rm km/s}$, $\tau_{a0} = R_{\odot} / v_{a0} = 356.8 \, {\rm s}$, as units
for length, density, magnetic field, velocity and time respectively.
Due to the large peak Alfv\'en speed ($\sim 12 v_{a0} \sim 24,000$ km/s) in
the domain (see Figure \ref{fig3}), we initially drive the emergence of the
twisted torus through the lower boundary at a fairly high speed over a period
of $t=0$ to $t=1.2$ with $v_{\rm rise} = 0.05 v_{a0} \sim 98 $km/s, which
is just under the sound speed at the lower boundary but significantly slower
than the Alfv\'en speed. In this way we build up the pre-eruption coronal
magnetic field approximately quasi-statically and yet fast enough to minimize
numerical diffusion.  After $t=1.2$, we significantly reduce the driving
speed of the flux emergence at the lower boundary to
$v_{\rm rise} = 0.01 v_{a0}$ and
thus allow the coronal magnetic field to evolve quasi-statically until it
erupts dynamically.

\section{Results\label{sec:result}}
Figures \ref{fig4} and \ref{fig5} show snapshots of the 3D coronal magnetic
field evolution (as viewed from 2 different perspectives) after the initial
stage of relatively fast emergence has ended at
$t=1.2$, and the speed for driving the flux emergence at the lower
boundary has been reduced to $v_{\rm rise} = 0.01 v_{a0}$.
The view shown in Figure \ref{fig4} corresponds to the observation
perspective at the time of the flare, for which the center of the emerging
region (also the center of the simulation lower boundary) is located at
$7.1^{\circ}$S and $24^{\circ}$W from the solar
disk center (or the line-of-sight).  GIF movies
for the evolution shown in Figure \ref{fig4} and Figure \ref{fig5} are
available in the electronic version of the paper.
We see that the emerged coronal flux rope
settles into a quasi-static rise phase and then
undergoes a dynamic eruption.
Figure \ref{fig6} shows the evolution of the rise velocity $v_r$ measured
at the apex of the tracked axis of the emerged flux rope (triangle points), and
also measured at the leading edge of the flux rope (crosses). After the
emergence is slowed down at $t=1.2$, the rise velocity at the apex of
the flux rope axis slows down, and undergoes some small oscillations as the
flux rope settles into a quasi-static rise. The quasi-static rise phase extends
from about $t=1.2$ until about $t=2.5$, over a time period of $1.3$, long
compared to the dynamic time scale of $\sim 0.1$ for the estimated Alfv\'en
crossing time of the flux rope.  At about $t=2.5$, the flux rope axis starts to
accelerate significantly and a dynamic eruption ensues. The flux emergence is
stopped at $t=2.8$, after which the flux rope continues to accelerate outward.
We are able to follow the acceleration of the axial field line up to
$v_r = 0.54 = 1050 $ km/s at $t=3.$, when the axial field line undergoes
a reconnection and we are subsequently unable to track it.
Figure \ref{fig6} also shows $v_r$ measured at the leading edge of the low
density cavity (as shown in Figure \ref{fig7}), corresponding to the
expanding flux rope.
We find that by the time of about $t=3.2$, a shock front followed by a
condensed sheath has formed ahead of the flux rope cavity
(see Figure \ref{fig7} at $t=3.25$),
and the $v_r$ measured at the front edge of the
cavity (or the inner edge of the sheath) reaches a steady speed of
about 0.425 or 830 km/s (see crosses in Figure \ref{fig6}).

When the flux rope begins significant acceleration (at $t \approx 2.5$),
the decay index $n \equiv d \ln | {\bf B}_p | / d \ln h$
which describes the rate of decline of the corresponding potential field
${\bf B}_p$ with height $h$ is found to
be $n \approx 1.2$ at the apex of the flux rope axis, and $n \approx 1.4$ at
the apex of the flux rope cavity.  These values are smaller than the critical
value of $n_{\rm crit} = 1.5$ for the onset of the torus instability for a
circular current ring \citep{bateman1978, kliem_toeroek2006,
demoulin_aulanier2010}, although there is a range of variability for
the critical value $n_{\rm crit}$, which can be as low as $1$, depending on
the shape of the current channel of the flux rope
\citep[e.g.][]{demoulin_aulanier2010}. For a 3D anchored flux rope, as is the
case here, it is difficult to obtain an analytical determination of
$n_{\rm crit}$ for the instability or loss of equilibrium of the flux rope
\citep{isen_forbes2007}.  The exact critical point for the onset of the torus
instability would depend on the detailed 3D magnetic field configuration.
On the other hand, a substantial amount of twist has been transported into the
corona at the onset of eruption. At $t=2.5$, the self-helicity of the emerged
flux rope reaches about $-1.02 \Phi_{\rm rope}^2$, where $\Phi_{\rm rope}$ is
the total magnetic flux in the rope, corresponding to field lines in the
flux rope winding about the central axis by about 1.02 rotations between the
anchored foot points.  This suggests the possible development of the helical
kink instability of the flux rope
\citep[e.g.][]{hood_priest1981, toeroek_kliem2005, fan_gibson2007}.
The erupting flux rope is found to undergo substantial writhing or
kinking motion as can be seen in the sequences of images
(also the movies in the electronic
version) in Figures \ref{fig4} and \ref{fig5}.

We also find that the trajectory for the eruption of the flux rope is not
radial because of the ambient coronal magnetic field: the erupting flux rope is
deflected southward and eastward from the local radial direction (see Figures
\ref{fig4}, \ref{fig5}, and \ref{fig7} and the associated movies).
Using the apex location of the erupting flux rope cavity at $t=3.25$
(Figure \ref{fig7}),
we find that the erupting trajectory at that time is deflected by
$2.3^{\circ}$ southward and $1.3^{\circ}$ eastward from the local radial
direction at the center of flux emergence, and further deflection of the
trajectory continues with time. Since the local radial direction at the
center of the flux emergence corresponds to $7.1^{\circ}$S and $24^{\circ}$W
from the solar disk center (or the line-of-sight), the deflection during the
eruption is sending the flux rope towards the line-of-sight in the east-west
direction, but further southward away from the line-of-sight in the
north-south direction.  This is consistent with the observed halo of the
CME seen in LASCO C2 and C3 coronagraphs
(Figure 2 in \citet{kataokaetal2009} and Figure 1 in
\citet{ravindra_howard2010}), where the north-south
and east-west asymmetries of the halo distribution indicate that the direction
of ejection is more southward and less westward than what would have been
expected for a radial ejection from the location of the source region on the
solar disk.

Figure \ref{fig8} shows the coronal magnetic field as viewed from
the side (panels a and b) and viewed from the observing perspective
(panels c and d) just before the onset of eruption at $t=2.45$, compared
with the Hinode XRT image of the region (panel e) just before the flare.
We see that the morphology of the
coronal magnetic field and its connectivity are very similar to those shown
in the X-ray image.  To understand the nature of the bright X-ray sigmoid
in the image, we have identified the region of significant magnetic energy
dissipation and heating in the simulated magnetic field using both the
electric current density $J \equiv | \nabla \times {\bf B} |$
and the increase of entropy
$\Delta S = C_v \Delta \ln (p/{\rho}^{\gamma} )$.
As pointed out in Section \ref{sec:model}, since we are solving the total 
energy equation in conservative form, numerical dissipation of magnetic
energy and kinetic energy due to the formation of current sheets and other
sharp gradients is being implicitly put back into the
thermal energy of the plasma, resulting in an increase of the entropy.
We have identified regions where there is significant entropy increase with
$\Delta S / C_v > 1.15$ and also high electric current density concentration
with $J/B > 1/l$ where $l = 10$ times the grid size.  
Such regions are outlined by the orange iso-surfaces in panels (a) and (c)
of Figure \ref{fig8}, and they appear as an inverse-S shaped layer (as viewed
from the top), which likely corresponds to the formation of an electric current
sheet underlying the anchored flux rope \citep[e.g][]{td1999,
low_berger2003,gibsonetal2006}.
We have also plotted field lines (purple field
lines shown in panels b and d) going through the region of the current
layer, which are preferentially heated and are expected to brighten
throughout their lengths (due to the high heat conduction along the field
lines) in soft-X ray, producing the central dominant X-ray sigmoid seen in
the Hinode XST image (panel e).  Thus our quasi-equilibrium coronal magnetic
field resulting from the emergence of a nearly east-west oriented magnetic
flux rope could reproduce the observed overall morphology and connectivity of
the coronal magnetic field, including the presence of the observed
pre-eruption X-ray sigmoid.
We find that both $J/B$ as well as $\Delta S$ peak
along the ``left elbow'' portion of the current layer, where the positive
polarity flux of the emerged flux rope comes in contact with the flux of
the dominant pre-existing negative polarity sunspot,
consistent with the brightness distribution along the
observed X-ray sigmoid (panel e of Figure \ref{fig8}). Reconnections in this
part of the current layer cause some of the flux in the emerged flux rope to
become connected with the major
negative sunspot (see the green field lines connecting
between the dominant negative spot and the emerging positive
spot in panel (d) of Figure \ref{fig8}).
We have also done a few simulations where we varied the tilt of the emerging
flux rope, and found that to reproduce the observed orientation of the sigmoid,
the emerging flux rope needs to be nearly east-west oriented.

With the onset of the eruptive flare, the soft-X ray observation
first shows a transient brightening of the sigmoid, and
subsequently the emission is completely dominated by the brightness of the
post-flare loops (see panels (a)(c)(e) of Figure \ref{fig9}).
In the simulated coronal magnetic field, we find that the current density in the
inverse-S shaped current layer intensifies as the flux rope begins to erupt.
We can deduce qualitatively the evolution of the post-reconnection
(or post-flare) loops from our modeled magnetic field evolution.
We traced field
lines (see the red field lines in panels (b)(e)(h) of Figure \ref{fig10}
and panels (b)(d)(f) of Figure \ref{fig11})
whose apexes are located in the layer of the most intense current density and 
heating. These field lines are the ones who have just reconnected at
their apexes and would slingshot downwards, corresponding to the downward
collapsing post-flare loops.  The layer of the most intense current density
and heating, as outlined by the orange iso-surfaces in
panels (a)(d)(g) of Figure \ref{fig10} and panels (a)(c)(e),
is identified as where $J/B > 1/l$
with $l = $ 5 times the grid size, and where $\Delta S / C_v > 2.3$.
This most intense current layer is found to rise upward with
the eruption of the flux rope. The associated
post-reconnection field lines are initially low lying and form a narrow
sigmoid shaped bundle as can be seen in Figures \ref{fig10}(b) and
\ref{fig11}(b).
With time, the post-reconnection loops broaden and rise up, showing cusped
apexes (Figures \ref{fig10}(e)(h) and Figures \ref{fig11}(d)(f)).
The morphology of the
post-reconnection loops, which transition from an initially narrow low-lying
sigmoid bundle to a broad, sigmoid-shaped row of loops with cusped apexes is
in qualitative agreement with the observed evolution of the post-flare
X-ray brightening shown in Figures \ref{fig9}(a)(c)(e).

The foot points of the post-reconnection loops (panels
(c)(f)(i) of Figures \ref{fig10}) can be compared qualitatively with
the evolution of the flare ribbons in the lower solar atmosphere as
shown in the Hinode SOT observation (panels (b)(d)(f) of
Figure \ref{fig9}).  The ribbon corresponding to the positive
polarity foot points (orange ribbon in Figures \ref{fig10}(c)(f)(i)) of
the post-reconnection loops is found
to sweep southward across the newly emerged positive polarity spot,
similar to the apparent movement of the positive polarity ribbon seen the
observation (panels (b)(d)(f) of Figure \ref{fig9}) in relation to the
observed positive emerging spot.
For the ribbon corresponding to the negative polarity footvpoints (the yellow
ribbon in Figures \ref{fig10}(c)(f)(i)), its eastern portion
is found to extend and sweep northward into the dominant pre-existing
negative spot, while
its western hook-shaped portion is found to sweep northward across the newly
emerged negative spot.  Similarly in the SOT observation (panels (b)(d)(f)
of Fig \ref{fig9}), for the
negative polarity ribbon, the eastern portion sweeps northward into the dominant
negative sunspot, while its western, upward curved hook-shaped portion is
found to sweep northward across the minor, fragmented negative pores which
have emerged to the west of the main $\delta$-sunspot.
The modeled ribbons based on the footpoints certainly differ in many ways
in their shape and extent compared to the observed flare ribbons.  But they
capture some key qualitative features in the observed motions of the flare
ribbons in relation to the photospheric magnetic flux concentrations.

\section{Discussions\label{sec:conc}}
\begin{figure}
\epsscale{0.8}
\plotone{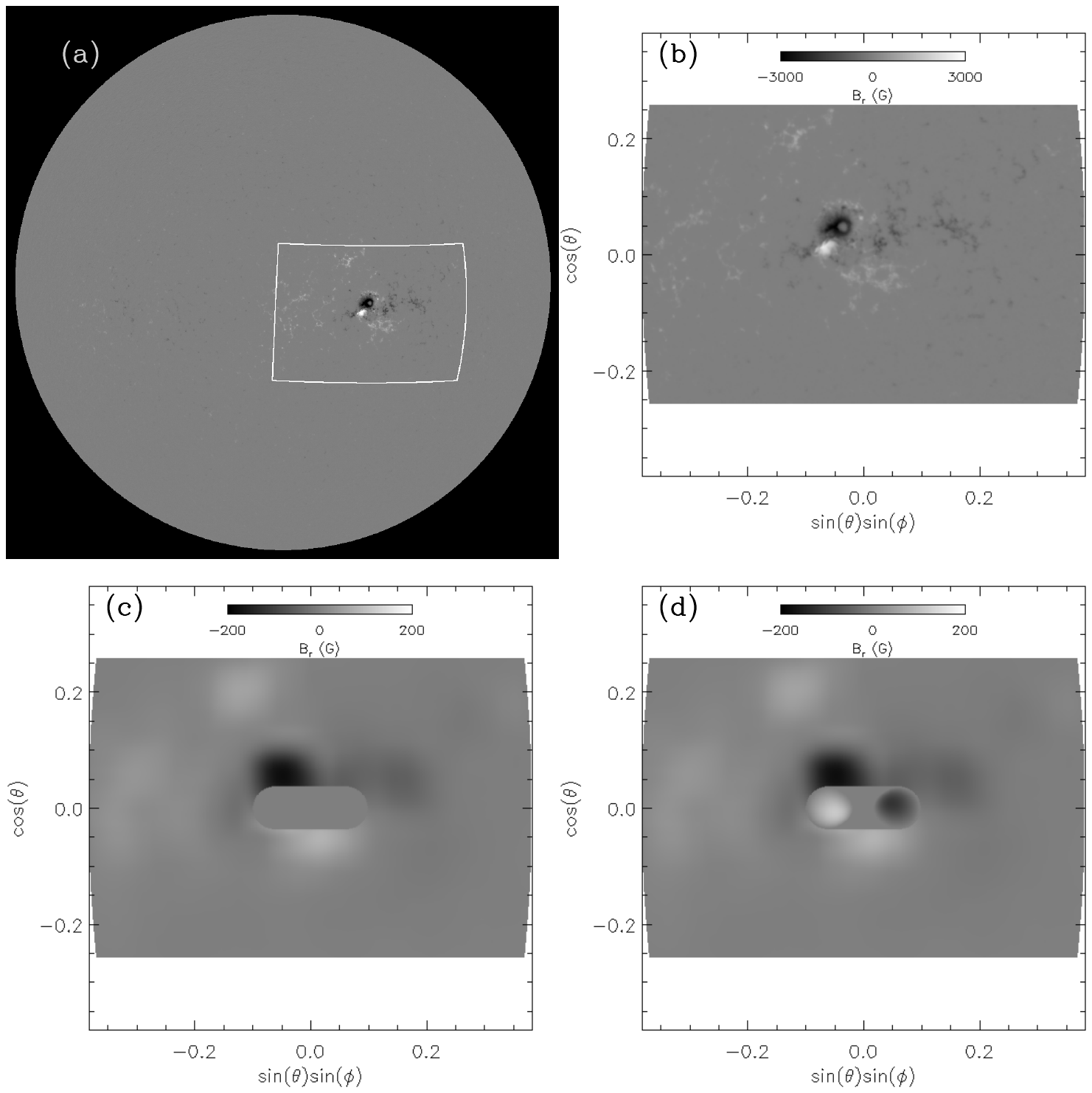}
\caption{(a) MDI full-disk magnetogram taken at 20:51:01 UT on December 12,
2006.  The surface area enclosed by the white box corresponds to the lower
boundary surface of the simulation domain. (b) $B_r$ on the lower boundary
region as viewed straight-on from the center of the region.
(c) $B_r$ on the lower
boundary after applying a Gaussian smoothing and with the field in a central
region being zeroed out for imposing the emergence of a twisted magnetic flux
rope. The potential field extrapolated from the $B_r$ shown here is the
assumed initial field in the simulation domain (see Fig. \ref{fig2}
below) . (d) $B_r$ on the lower boundary at the end of emergence of the
twisted flux rope.}
\label{fig1}
\end{figure}

\begin{figure}
\epsscale{1.}
\plotone{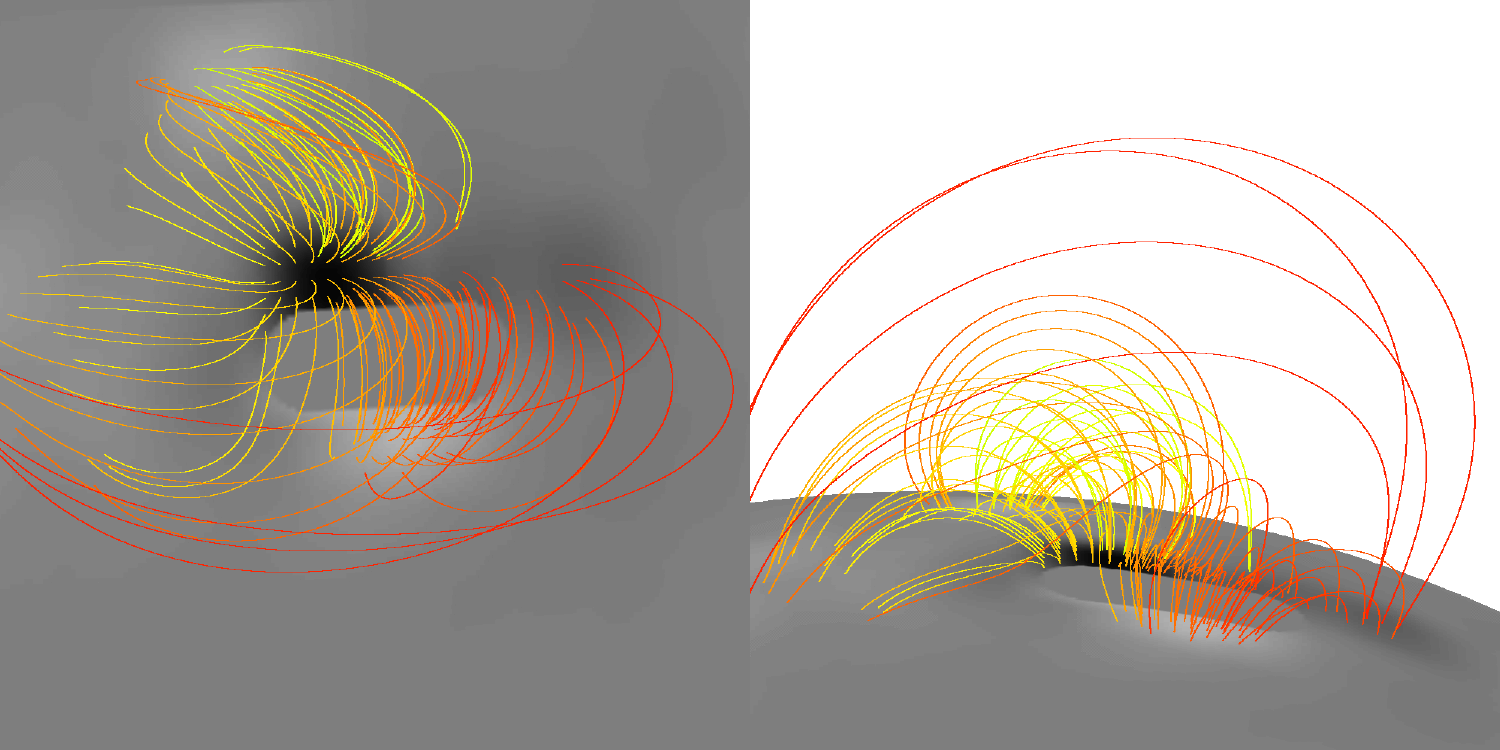}
\caption{Selected field lines of the initial potential magnetic field for the
simulation as viewed from two different perspectives.}
\label{fig2}
\end{figure}

\begin{figure}
\epsscale{1.}
\plotone{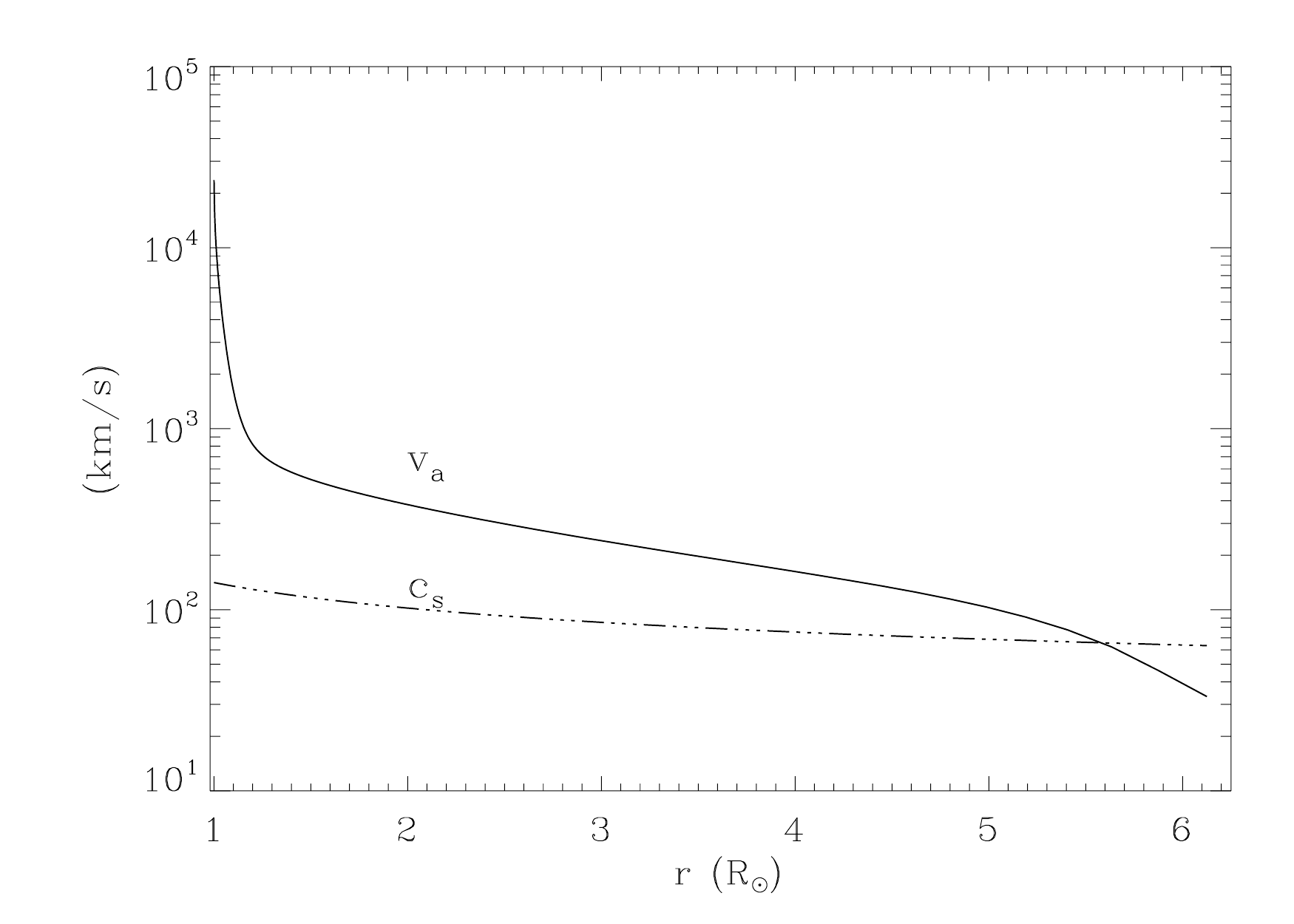}
\caption{The Alfv\'en speed and the sound speed as a function of raidial
distance along a vertical line rooted in the peak $B_r$ of the main
pre-existing negative spot on the lower boundary}
\label{fig3}
\end{figure}

\begin{figure}
\epsscale{0.6}
\plotone{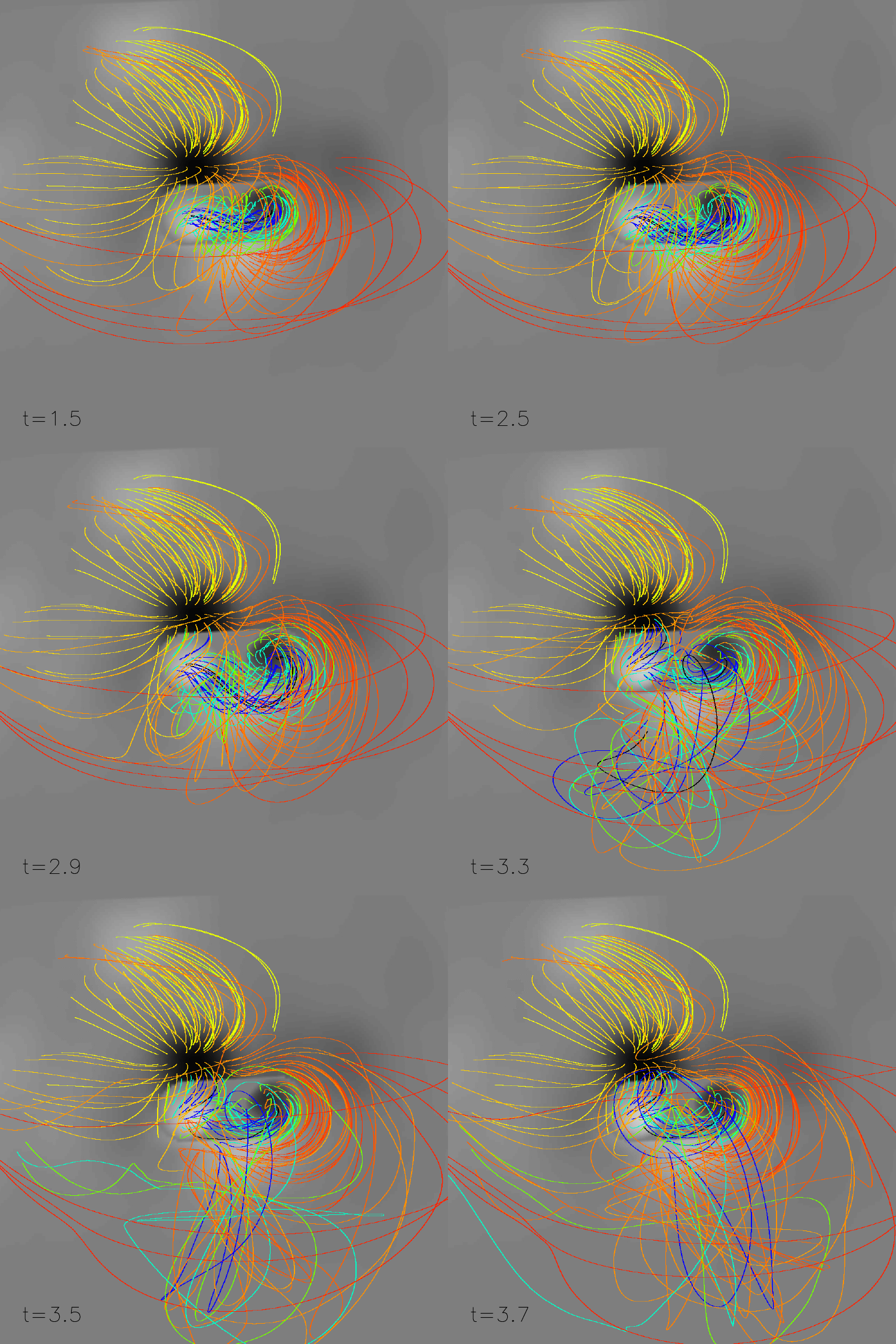}
\caption{Snapshots of the 3D coronal magnetic
field evolution after $t=1.2$ (when the initial fast emergence phase has
ended), showing the quasi-static rise and then the eruption of the coronal
flux rope. The field is viewed from the observation perspective at the time of
the observed flare, where the center of the emerging flux rope is
located at $7.1^{\circ}$S and $24^{\circ}$W from the solar disk center
(or the line-of-sight). A .gif movie of the above evolution is available in the
electronic version of the paper}
\label{fig4}
\end{figure}

\begin{figure}
\epsscale{0.6}
\plotone{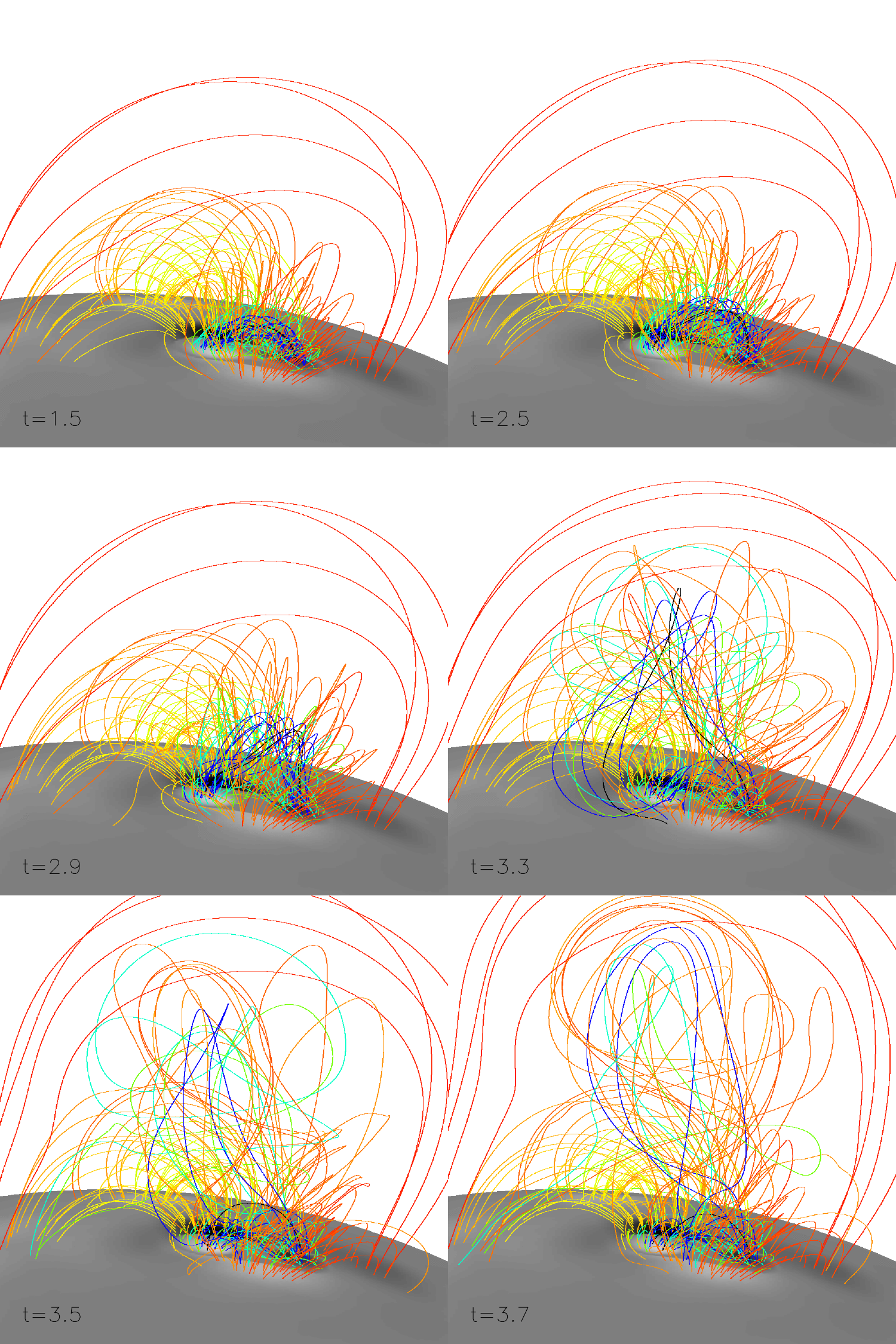}
\caption{Same as Figure \ref{fig4} except viewed from a different
perspective.  A .gif movie of the evolution is available in the electronic
version of the paper}
\label{fig5}
\end{figure}

\begin{figure}
\epsscale{1.}
\plotone{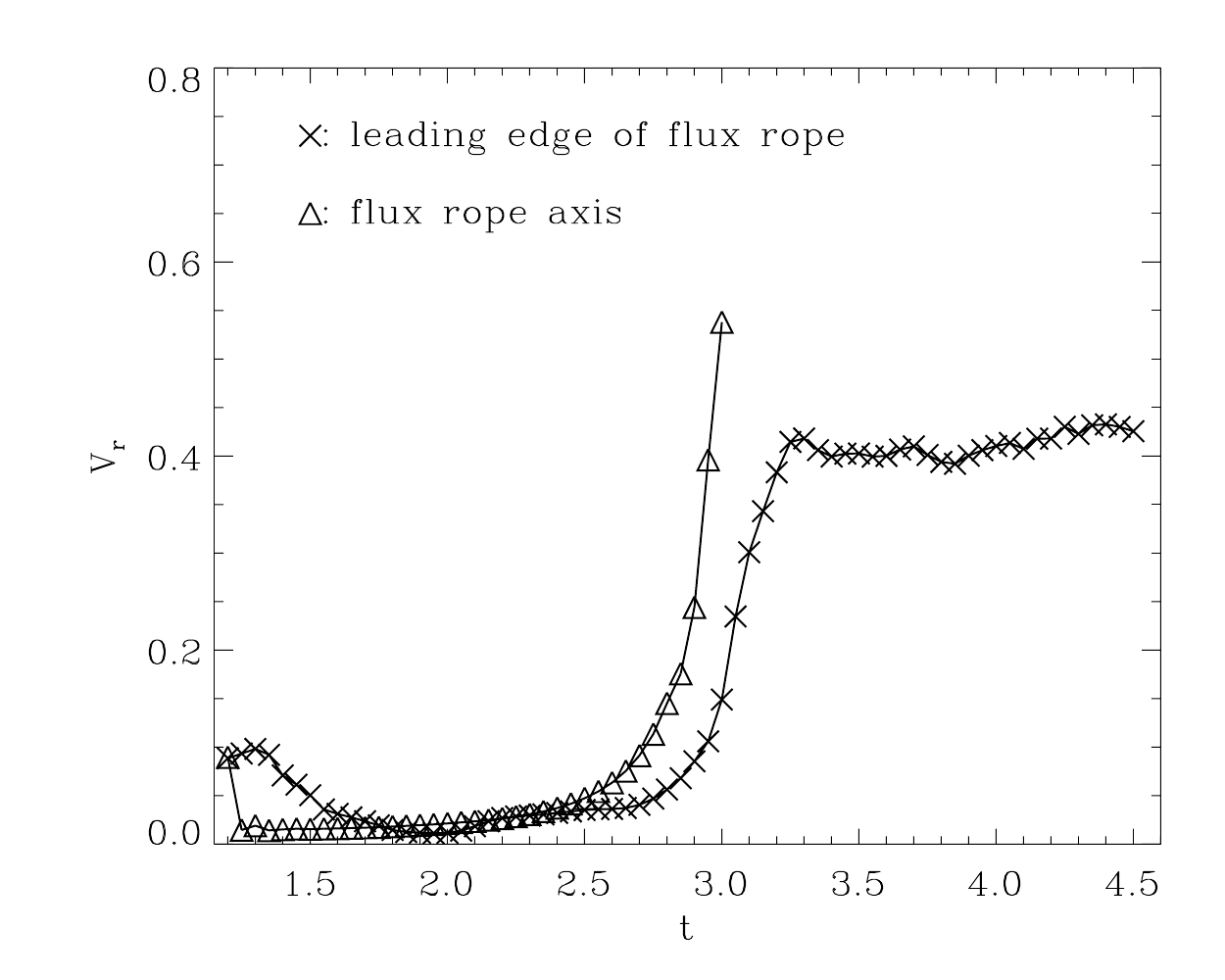}
\caption{The evolution of the radial velocity measured at the apex of the
tracked axial field line of the emerged flux rope (triangles) and also
measured at the leading edge of the flux rope cavity (crosses).}
\label{fig6}
\end{figure}

\begin{figure}
\epsscale{0.6}
\plotone{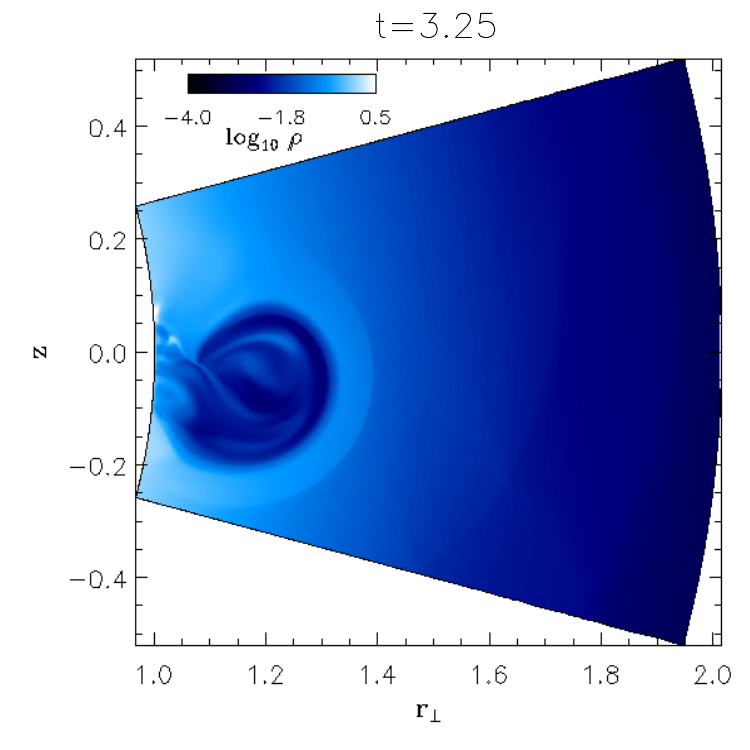}
\caption{A meridional slice of density through the middle of the erupting
flux rope. It shows a low density cavity corresponding to the expanding
flux rope. A shock front has formed with a dense sheath compressed between
the flux rope and the shock front.}
\label{fig7}
\end{figure}

\begin{figure}
\epsscale{0.6}
\plotone{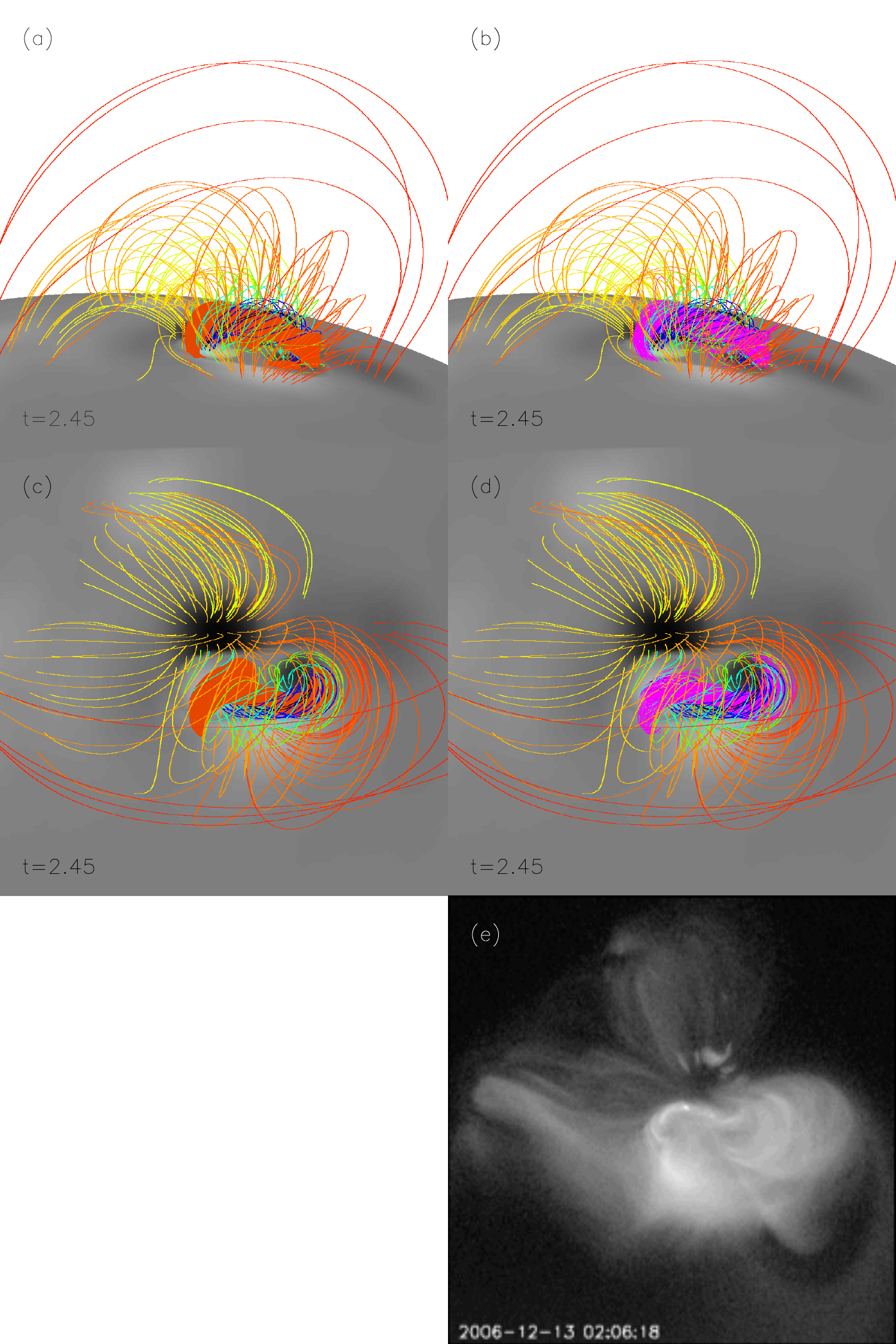}
\caption{Coronal magnetic field as viewed from the side (a and b) and
viewed from the observing perspective (c and d) just before the onset
of eruption at $t=2.45$, compared with the Hinode XRT image of the region (e)
just before the flare. The orange surfaces are the iso-surfaces where
$J/B = 1/l$ with $l=$ 10 times the grid size and where $\Delta S / C_v > 1.15$.
They outline the region of
strong electric current layers. The purple field lines are the field lines
that go through the points in the current layer.}
\label{fig8}
\end{figure}

\begin{figure}
\epsscale{0.8}
\plotone{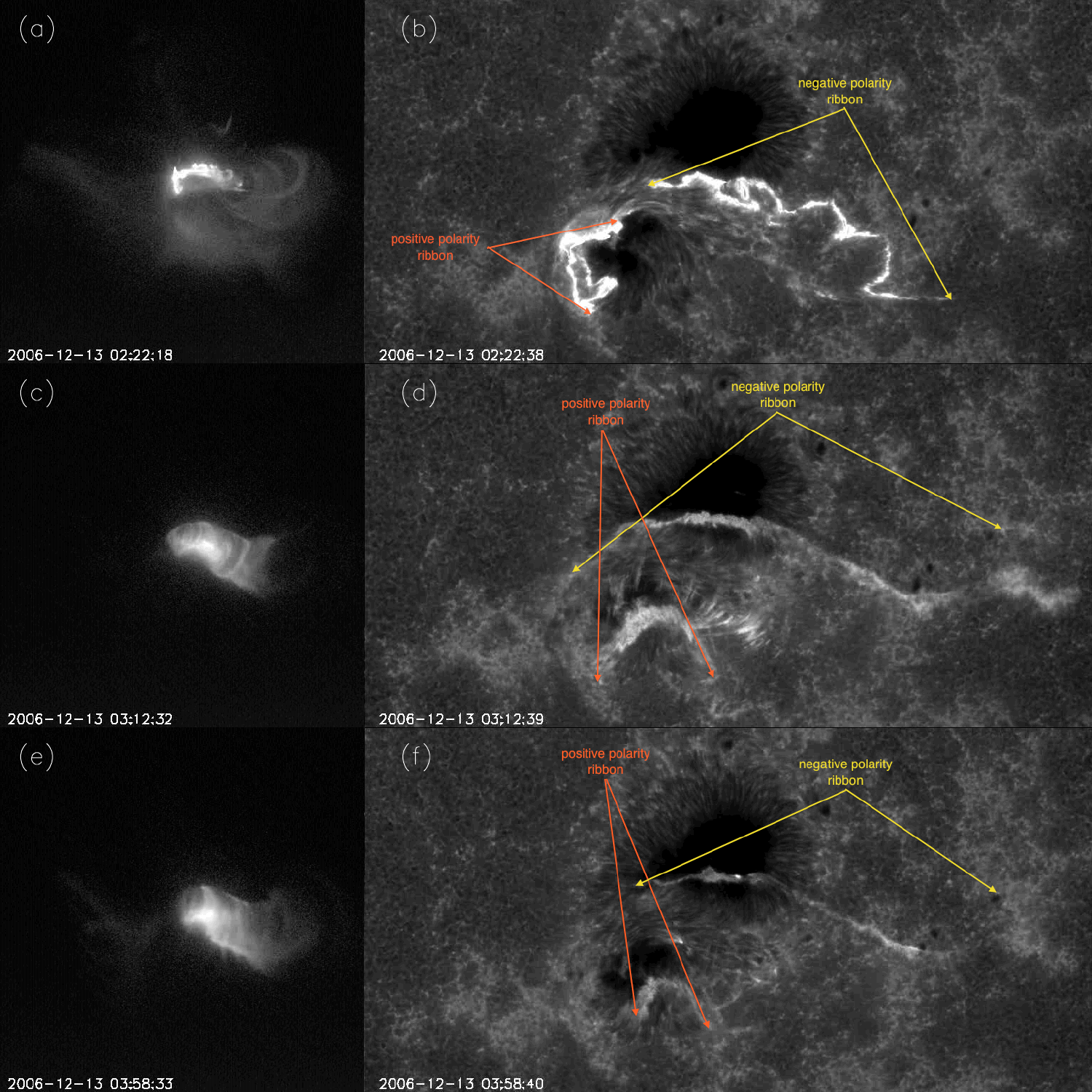}
\caption{Hinode XRT images of the post-flare brightening
(left column images), and the corresponding Hinode SOT snapshots in Ca II line
showing the chromosphere flare ribbons (right column images). The orange
(yellow) arrows indicate the extent of the flare ribbon in the positive
(negative) polarity.}
\label{fig9}
\end{figure}

\begin{figure}
\epsscale{0.8}
\plotone{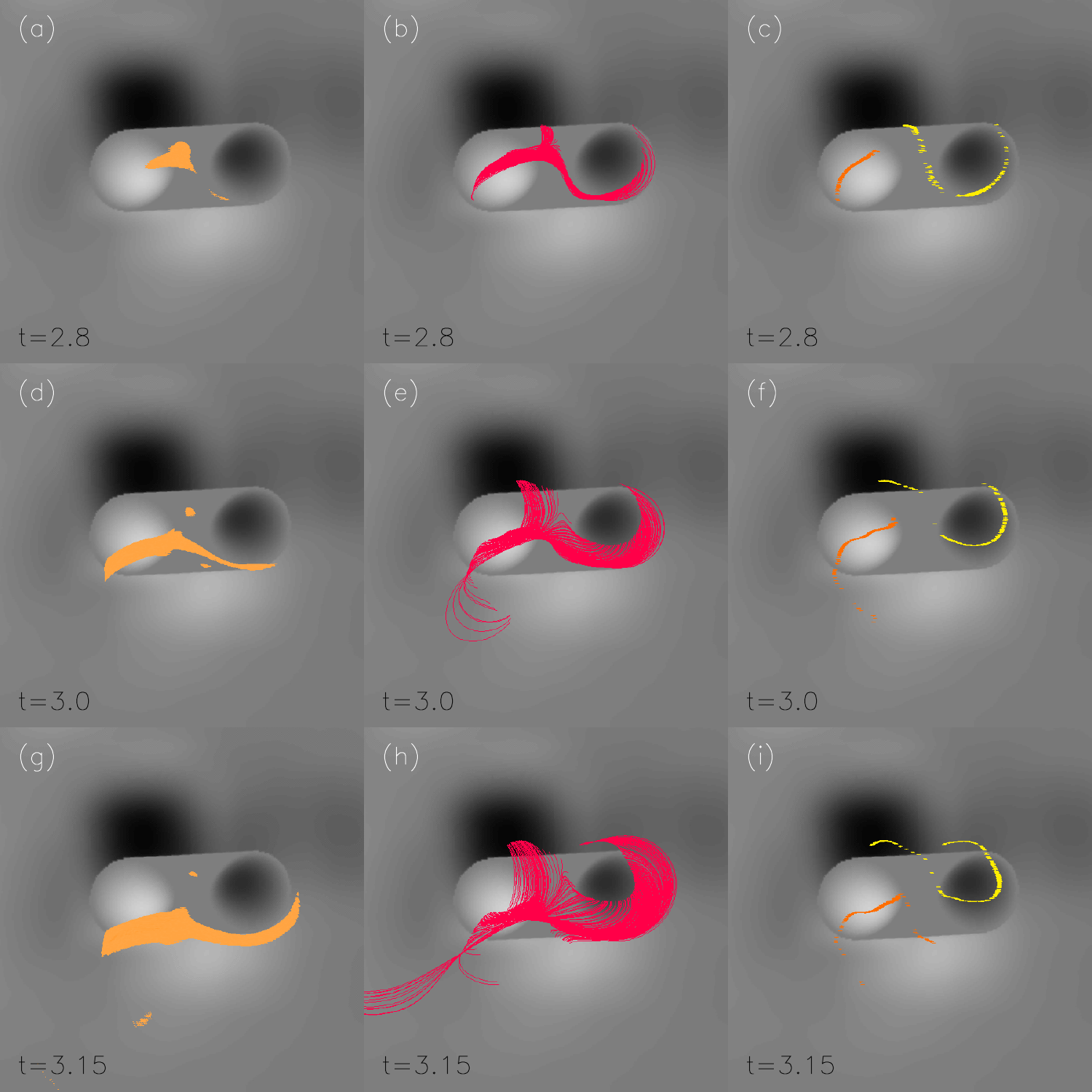}
\caption{Left column images: iso-surfaces where $J/B = 1/l$ with $l = $ 5
times the grid size and where $\Delta S / C_v > 2.3$, outlining the
the most intensely heated portion of the current layer. Middle column images:
sampled field lines whose apexes are in the intense current
layer outlined by iso-surfaces in the left column images,
corresponding to the post-reconnection loops.
Right column images: foot points of the
post-flare loops shown in the middle column images.  The gray scale images
in all panels show the normal magnetic field at the lower boundary of the
simulation domain}
\label{fig10}
\end{figure}

\begin{figure}
\epsscale{0.8}
\plotone{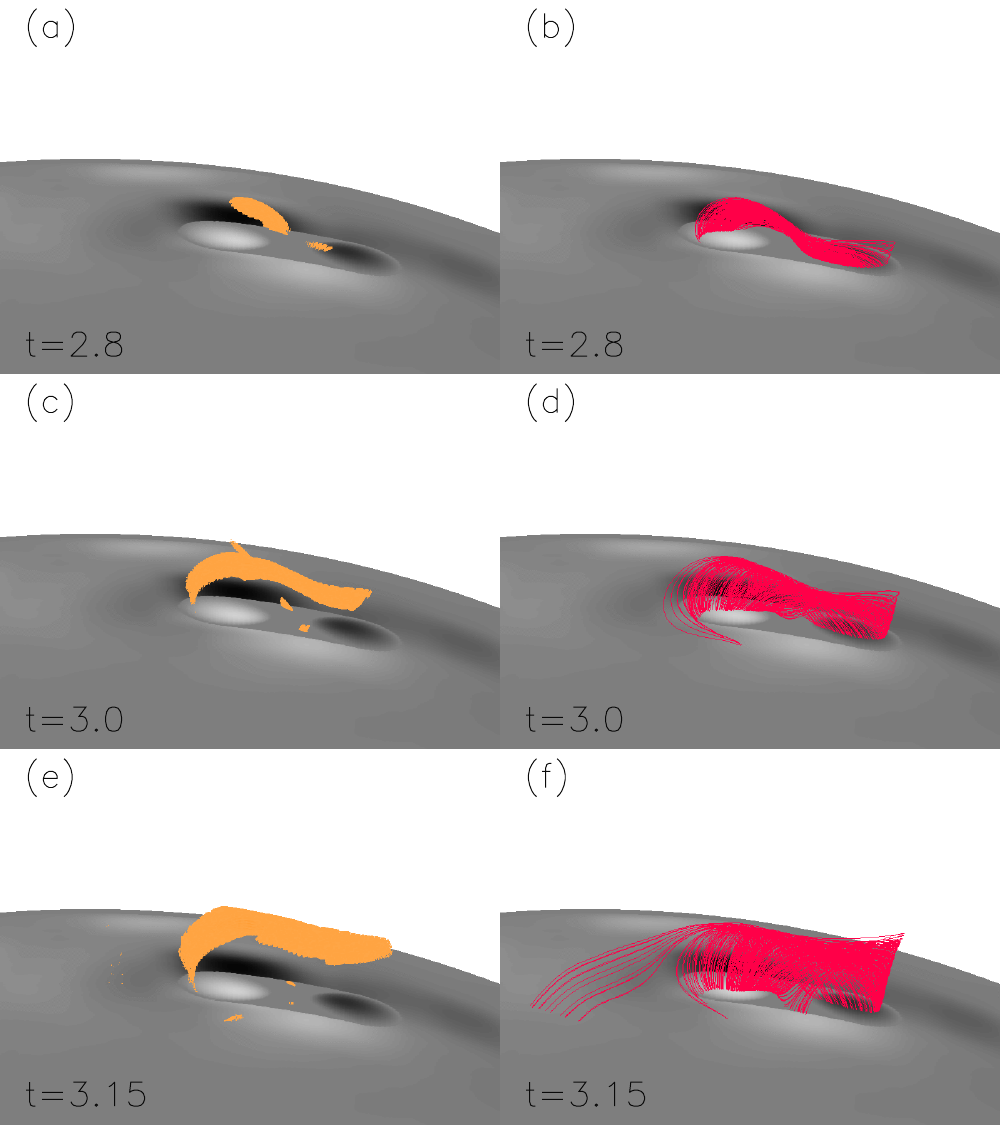}
\caption{Same as the left and middle columns of Figure \ref{fig10} but
viewed from a different perspective.}
\label{fig11}
\end{figure}

We have presented an MHD model that qualitatively describes the coronal
magnetic field evolution of the eruptive flare in AR 10930 on December 13,
2006.  The model assumes the emergence of an east-west oriented magnetic
flux rope into a pre-existing
coronal magnetic field constructed based on the MDI full-disk magnetogram
of the photospheric magnetic field at 20:51:01 UT on December 12.
As described in Section \ref{sec:model}, a substantial
smoothing of the observed photospheric magnetic flux density from the MDI
magnetogram is carried out such that the peak field strength on the lower
boundary is reduced from $\sim 3000$ G to $\sim 200$ G to avoid the
extremely high Alfv\'en speed that would put too severe a limit on the time
step of numerical integration.
The imposed flux emergence at the lower boundary of
an idealized subsurface magnetic torus
produces a flux emergence pattern on the lower
boundary that is only qualitatively representative of the observed flux
emergence pattern (compare Figure \ref{fig1}b and Figure \ref{fig1}d).
In the model, the emerging bipolar pair on the lower boundary is more
symmetric, more spread-out in spatial extent, and both polarities are
transporting left-handed twist (or injecting negative helicity flux) into
the corona at the same rate.
Whereas in the observation, the positive emerging sunspot is coherent and
clearly shows a counter-clockwise twisting motion, indicating an injection of
negative helicity flux into the corona, while its counterpart to the west is
in the form of fragmented pores \citep[e.g.][]{min_chae2009}. However a
quantitative measurement by \citet{parketal2010} using MDI magnetograms
also found a significant negative
helicity flux associated with these fragmented pores as well (see Figure 4 in
their paper).
In the simulation, the self-helicity of the emerged portion of the flux rope
in the corona at the end of the imposed flux emergence (at $t=2.8$)
is $H_{\rm rope} \approx 1.07 \Phi^2$,
where $\Phi$ is the normal flux in each
polarity of the emerged bipolar region on the lower boundary.
This is a measure of the internal twist in the emerged flux
rope and it corresponds to field lines twisting about the axis by about 1.07 winds
(or $385^{\circ}$ rotation) between the two anchored ends
in the emerged flux rope.  On the other hand, the total relative magnetic
helicity $H_{\rm tot}$ that has been transported into the corona by the
imposed flux emergence is found to be $H_{\rm tot} \approx 3.02 \Phi^2$, which
is the sum of both the self-helicity of the emerged flux rope $H_{\rm rope}$
as well as the mutual helicity between the emerged flux and
the pre-existing coronal magnetic field.
The observed amount of rotation of the positive emerging sunspot,
ranging from $240^{\circ}$ \citep{zhangetal2007}
to $540^{\circ}$ \citep{min_chae2009},
gives an estimate of $( H_{\rm rope}/ \Phi ^2 ) \times 360^{\circ}$ for the
emerged flux rope, which is about $385^{\circ}$ in the simulation and is
thus within the range of the observed values.

After an initial phase where we drive the emergence of the twisted torus at a
fairly large (but still significantly sub-Alfv\'enic) speed to
quickly build up the pre-eruption field, we slow down
the emergence and the coronal magnetic field settles into a quasi-equilibrium
phase, during which the coronal flux rope rises quasi-statically as more
twist is being transported {\it slowly} into the corona through continued
flux emergence.  This phase is followed by a dynamic eruption phase where
the coronal flux rope accelerates in the dynamic time scale to a steady
speed of about 830 km/s.
Due to the substantial twist (greater than 1 full wind of field line twist)
that has been transported into the corona at the onset of the eruption, the
erupting flux rope is found to undergo substantial writhing motion. The
erupting flux rope underwent a counter-clockwise rotation that exceeded
$90^{\circ}$ by the time the front of
the flux rope cavity reached $1.4 R_{\odot}$. We also find that the
initial trajectory of the erupting flux rope is not radial, but is deflected
southward and eastward from the local radial direction due to
the ambient coronal magnetic field. Since the initial coronal flux rope is
located at $7.1^{\circ}$S and $24^{\circ}$W from the solar disk center,
the deflection is sending the erupting flux rope towards
the line-of-sight in the east-west direction, but further away from the
line-of-sight in the north-south direction,  consistent with the
observed halo of the CME seen in LASCO C2 and C3 coronagraphs,
where the halo's north-south (east-west) asymmetry appears larger
(smaller) than would have been expected from a radial eruption of the flux
rope from its location on the solar disk.
However, due to the relatively restrictive domain width in $\theta$
($30^{\circ}$) and $\phi$ ($45^{\circ}$) in our current simulation, the
side wall boundary in the south begins to significantly constrain the
further southward deflection and expansion of the flux rope by the time the
top of the flux rope cavity reaches about $1.4 R_{\odot}$.  Thus, we are
not able to accurately determine the subsequent trajectory change or 
the continued writhing of the erupting flux rope beyond this point.
A larger simulation with a significantly greater domain size in $\theta$ and
$\phi$, that still adequately resolves the coronal magnetic field in the 
source region, will be carried out in a subsequent study to determine
the later properties of the flux rope ejecta.

The restrictive domain size may also play a role in the significantly
lower steady speed of $830$ km/s reached by the erupting flux rope in the
simulation, compared to the observed value of at least $1780$ km/s for the
speed of the CME \citep[e.g.][]{ravindra_howard2010}.
It has been shown that the rate of spatial decline of the ambient potential
magnetic field with height is both a critical condition for the onset of the
torus instability of the coronal flux rope \citep[e.g.]{bateman1978,
kliem_toeroek2006, isen_forbes2007, fan2010, demoulin_aulanier2010} as well as
an important factor in determining the acceleration and the final speed of
the CMEs \citep{toeroek_kliem2007}. Even for a sufficiently twisted coronal
flux rope that is unstable to the helical kink instability, the spatial
decline rate of the ambient potential field is found to determine whether the
non-linear evolution of the kink instability leads to a confined
eruption (with the flux rope settles into a new kinked equilibrium) or an
ejection of the flux rope \citep{toeroek_kliem2005}. The simulation in this
paper has assumed perfect conducting walls for the side boundaries where
the field lines are parallel to the walls.  Thus widening the simulation domain
would result in a more rapid expansion and hence a more steep decline of the
ambient potential field with height. This would result in a greater
acceleration and a faster final speed for the CME based on the results from
previous investigations by \citet{toeroek_kliem2005,toeroek_kliem2007}. It may
be difficult to distinguish whether the torus or the kink instability
initially triggers the eruption given the complex 3D coronal magnetic field,
but the final speed of the CME would be strongly affected by the spatial
decline rate of the ambient potential field for either cases.
The substantial smoothing of the lower boundary magnetic field to reduce
the peak Alfv\'en speed is also a major reason for the low final speed
of the erupting flux rope in the current simulation.

Nevertheless, our simulated coronal magnetic field evolution is found
to reproduce several key features of the eruptive flare observed by Hinode.
The pre-eruption coronal field during the quasi-static phase reproduces the
observed overall morphology and connectivity of the coronal magnetic field,
including the presence of the pre-eruption X-ray sigmoid
seen in the Hinode XRT images.  The presence of the pre-eruption
sigmoid in our model is caused by the preferential heating of an
inverse-S shaped flux bundle in the flux rope by the formation of 
an inverse-S shaped current sheet underlying the flux rope.
Our simulations suggest that the emerging flux rope needs to be nearly
east-west oriented in order to reproduce the observed orientation of the
X-ray sigmoid.  This is consistent with the suggestion by \citep{min_chae2009} 
that the counterpart of the emerging, rotating positive sunspot is the
minor negative pores to the west of the emerging sunspot (rather than the
dominant negative sunspot).
After the onset of the eruption, the morphology of the post-flare loops
deduced from the simulated field show a transition from an initial narrow,
low-lying sigmoid bundle to a broad, sigmoid-shaped row of loops with cusped
apexes, in qualitative agreement with the evolution of the
post-flare X-ray brightening observed by XRT of Hinode.
The apparent motions of the foot points of the post-flare loops
in relation to the lower boundary magnetic flux concentrations are also
in qualitative agreement with the evolution of the chromospheric flare
ribbons observed by Hinode SOT.
These agreements suggest that our simulated coronal magnetic field produced by
the emergence of an east-west oriented twisted flux rope, with the positive
emerging flux ``butting against'' the southern edge of the
dominant pre-existing negative sunspot, captures the gross structure of the
actual magnetic field evolution associated with the eruptive flare.
To improve quantitative agreement, a more accurate determination of the
lower boundary electric field \citep{fisheretal2011} that more
closely reproduces the observed flux emergence pattern on the lower boundary
is needed.

\acknowledgements
I thank Laural Rachmeler for reviewing the manuscript and for helpful
comments.  NCAR is sponsored by the National Science Foundation.
This work is supported in part by the NASA LWS TR\&T grant NNX09AJ89G
to NCAR.
The numerical simulations were carried out on the Pleiades
supercomputer at the NASA Advanced Supercomputing Division under
project GID s0969.
Hinode is a Japanese mission developed and launched by ISAS/JAXA,
with NAOJ as domestic partner and NASA and STFC (UK) as international
partners. It is operated by these agencies in co-operation with ESA and
NSC (Norway).

\clearpage

\clearpage

\clearpage

\clearpage

\clearpage

\clearpage

\clearpage

\clearpage

\clearpage

\clearpage

\clearpage

\end{document}